\documentstyle[prl,aps,twocolumn,floats]{revtex}
\input psfig

\newcommand{\bleq}{\ifpreprintsty
                   \else
                   \end{multicols}\vspace*{-3.5ex}{\tiny
                   \noindent\begin{tabular}[t]{c|}
                   \parbox{0.493\hsize}{~} \\ \hline \end{tabular}}
                   \fi}
\newcommand{\eleq}{\ifpreprintsty
                   \else
                   {\tiny\hspace*{\fill}\begin{tabular}[t]{|c}\hline
                    \parbox{0.49\hsize}{~} \\
                    \end{tabular}}\vspace*{-2.5ex}\begin{multicols}{2}
                    \fi}
\newcommand{\bcols}{\ifpreprintsty\else\begin{multicols}{2}\fi}
\newcommand{\ecols}{\ifpreprintsty\else\end{multicols}\fi}

\def\mathcal{\cal}
\def \be{\begin{equation}}
\def \ee{\end{equation}}
\def \bmlett{\begin{mathletters}}
\def \emlett{\end{mathletters}}

\def \ve{\varepsilon}

\def \HH{{\mathcal H}}
\def \NN{{\mathcal N}}

\def \CC{{\mathcal C}}

\def \pd{\phantom{\dagger}}

\def \ua{\uparrow}
\def \da{\downarrow}
\def \ra{\rightarrow}

\def \sgn{{\rm sgn}}

\def \ndot{n_{\rm dot}}

\begin{document}

\bibliographystyle{simpl1}


\title{Interaction Induced Restoration of Phase Coherence}

\author{A. A. Clerk, P. W. Brouwer and V. Ambegaokar}
\address{Laboratory of Atomic and Solid State Physics,
Cornell University, Ithaca NY 14853, USA
\\
{\rm (September 5, 2001)}
\medskip ~~\\ \parbox{14cm}{\rm
We study the conductance of a 
quantum ``T-junction'' coupled to
two electron reservoirs and a quantum dot. In the absence of
electron-electron interactions, the conductance $g$
is sensitive to interference between trajectories which enter the
dot and those which bypass it.  We show that including an
intra-dot charging interaction has a marked influence-- it can
enforce a coherent response from the dot at temperatures much
larger than the single particle level spacing $\Delta$.  The result is
large oscillations of $g$ as a function 
of the voltage applied to a gate capacitively coupled to the dot.  Without
interactions, the conductance has only a weak interference signature when
$T>\Delta$.
}} \maketitle
How do interactions affect the phase coherence of electrons
traveling through a quantum dot? This question is of interest both
because of the fundamental issues it raises and because of
its relevance to recent experiments.  Groundbreaking
experiments in which a Coulomb blockaded quantum dot is embedded
in an Aharanov-Bohm ring \cite{YacobyEtAl} have demonstrated that
transport through such dots is at least partially coherent,
despite strong interactions.  Complimentary studies which
observed Fano resonances in the conductance of such dots also
substantiate this conclusion \cite{Gores}, while the
observation of weak localization and conductance fluctuations
has demonstrated coherence in open
(i.e., not Coulomb blockaded) dots \cite{MarcusReview}. In general,
electron-electron interactions are expected  
to degrade the coherence of
transport through a dot-- interference phenomena, such as the
amplitude of Aharanov-Bohm oscillations in the conductance, are 
suppressed compared to the non-interacting case \cite{ImryBook}.  
In this Letter we study a system in which the opposite phenomenon
occurs-- the presence of a charging interaction in a quantum dot
significantly enhances interference phenomena compared to the
situation without interactions \cite{BrouwerAleiner}.

\begin{figure}[b]
\centerline{\psfig{figure=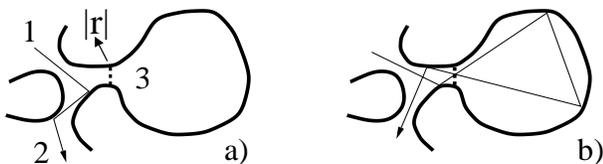,width=8.0cm}} \narrowtext
\caption{a) Schematic of the T-junction plus dot system, showing a direct
trajectory; such scattering events are described 
by the phase $\delta$.  
The effective tunnel junction in the entrance to the dot has
a reflection probability $|r|^2$.  b)  
Same, showing a scattering event which involves the dot.} 
\label{stubfig}
\end{figure}

Motivated by the experiments mentioned above, we consider a
T-junction system which is sensitive to the coherence of electrons
{\it reflected} from a quantum dot.  The T-junction consists
of three coincident single-mode quantum point contacts
coupled to source and drain reservoirs and to a quantum dot (Fig.
\ref{stubfig}). Without interactions,
the source-drain conductance $g$ is sensitive to constructive or destructive
interference between trajectories which bypass the dot, and those
which enter it. If electron motion in the dot is fully coherent, 
$g$ (in units of $e^2 / h$) is \cite{AashThesis}:

\begin{equation} \label{NonIntG}
    g = g_{\rm max} 
	\int d \ve \left(- \frac{df}{d\ve} \right) 
	\sin^2 \left( \delta + \alpha(\ve) \right).
\end{equation}

{\noindent}Here,  $\delta$ is the transmission phase shift
associated with direct trajectories bypassing the dot
(Fig. \ref{stubfig}a), 
$\alpha(\ve)$ is the phase shift for scattering from 
the dot through an effective tunnel junction with
reflection probability $|r|^2$ (Fig \ref{stubfig}b)\cite{AlphaDefn}, 
and $f$ is the
Fermi function.  
The parameters $\delta$, $|r|$ and $g_{\rm max}$
are determined by the $3 \times 3$ scattering matrix $S$ of the T-junction
without dot \cite{SDefn}.  This matrix 
only changes
appreciably over energies comparable to the Fermi energy $E_F$, and
can be treated as constant over the smaller scales we focus on.  
At zero temperature, the conductance (\ref{NonIntG})
exhibits full constructive and destructive interference as a
function of the phases $\delta$ and $\alpha$, resulting in
oscillations of $g$ between $0$ and its maximum value 
$g_{\rm max}$. While
$\delta$ is a property of the T-junction and cannot be tuned,
$\alpha$ depends on the dot and can be varied, e.g., by changing
$E_F$.  The typical energy scale for variations of $\alpha$ is
$\Delta$, the single-particle level spacing of the dot.

In this Letter, we focus on the temperature regime $T \gg \Delta$.
Here, the signatures of interference in Eq. (\ref{NonIntG}) are
washed out by thermal smearing, resulting in 
$g = \frac{1}{2} g_{\rm max} \left(1 - |r|
\cos 2 \delta \right)$.  We ask how this now
changes when the effects of intra-dot
electron-electron interactions are included.  We consider the capacitive
interaction 

\begin{equation} \label{HCDefn}
H_C = E_C (\ndot - \NN)^2, 
\end{equation}

{\noindent}where $E_C \ll E_F$ is
the charging energy, $\ndot$ the electron number on the dot, and
$\NN$ the dimensionless voltage of a gate electrode capacitively
coupled to the dot.  Although we consider $T \gg \Delta$, we also
require $T \ll E_C$, so that charged dot excitations are
suppressed.  
Our main result is that in this
regime, transport through the T-junction is more coherent
with interactions than without-- 
electrons scatter from the dot
with a well defined $\NN$-dependent phase, resulting in
interference and hence an $\NN$ dependent conductance.  
The origin of the resulting oscillations is
entirely different from that of 
standard Coulomb blockade oscillations, as there is
no ``blockade'' here-- electrons traveling from source to drain
are not forced to pass through the
dot. We find this result remarkable, as one usually expects that
electron-electron interactions degrade, rather than
enhance, coherence, due to the possibility of creating low-energy
particle-hole excitations in the dot.

The underlying reason for this enhanced coherence is a subtle
consequence of the charging interaction first discussed by Matveev
\cite{MatveevKondo}. Using the convention
\begin{equation} \label{NDefn}
\NN = N_0 + 1/2 + x  , \hspace{0.5 cm} |x| \leq 1/2,
\end{equation}
where $N_0$ is an integer, for $T \ll E_C$ only the states
$\ndot = N_0$ and $\ndot = N_0 + 1$ are dynamically
significant.  These states may be regarded as being the $\da$ and
$\ua$ states, respectively, of a fictitious impurity spin.  If we also 
assign lead (dot) electrons a fictitious $\ua$ ($\da$)
spin, then the tunneling Hamiltonian between lead and dot takes
the form of spin-flip scattering off an impurity.  In this way,
the Coulomb blockade problem can be mapped onto an
anisotropic Kondo model; for spinless electrons \cite{SpinlessNote}
this is a
single-channel Kondo (1CK) model, while with spin it is
a two-channel Kondo (2CK) model.  In this analogy, $|t|$
plays the role of the dimensionless exchange constant $J \rho_0$,
and $x$ the role of a local impurity magnetic field.  The coherent
reflection from the dot results from this effective
Kondo physics, as we will now discuss.

For a quantitative description, we write the Hamiltonian of the
T-junction and dot as $H = H_D + H_L + H_S + H_C$, where $H_D =
\sum_{\alpha \beta,\sigma} \HH_{\alpha \beta} d^{\dag}_{\alpha
\sigma} d^{\pd}_{\beta \sigma}$ is the Hamiltonian of the closed
dot, $H_L = \sum_{j=1,2} \sum_{\sigma} \int dk  \ve(k)
\psi^{\dag}_{j \sigma}(k) \psi^{\pd}_{j \sigma}(k)$ is the kinetic
energy of electrons in leads $1$ and $2$,
and $H_C$ is given in Eq. (\ref{HCDefn}).  Scattering in the
T-junction is described by
\begin{eqnarray}
H_S & = &  \sum_{\sigma,i=1,2} \int dk \Bigg[
    \sum_{j=1}^2 \int dk'  W_{ij}
    \psi^{\dag}_{i \sigma}(k) \psi^{\pd}_{j \sigma}(k')  \\
    &+& \nonumber \sum_{\alpha} \left(
    W_{i3} \psi^{\dag}_{i \sigma}(k) d^{\pd}_{\alpha \sigma}
    + h.c. \right) \Bigg]
    + \sum_{\sigma,\alpha,\beta}
    W_{33} d^{\dag}_{\beta \sigma} d^{\pd}_{\alpha \sigma},
\end{eqnarray}
{\noindent}where the $3 \times 3$ Hermitian matrix $W$ describes a
potential corresponding to the scattering matrix $S(E_F)$.
The dot electron number $\ndot$ in Eq. (\ref{HCDefn}) reads 
$\ndot = \sum d^{\dag}_{\alpha \sigma} d^{\pd}_{\alpha \sigma}$.

We start with the case of a weakly coupled dot ($|t|^2 \ll 1$). 
Using an approach similar to that of Ref.\cite{Wingreen}, 
we express the source-drain
conductance
in terms of the single-particle retarded Green
function $G^R_{\alpha \beta}(\omega)$ of the dot. As
the dot is coupled to only a single point contact, and as
$G^R$ is diagonal in spin, one can obtain an exact
expression involving {\em only} $G^R$ evaluated at the contact
\cite{AashThesis}:
\begin{equation}
\frac{g}{g_{\rm max}} = 
	\sin^2 \delta +
	\frac{\Gamma}{4} \text{Im } e^{2 i \delta} \sum_\sigma 
	\int d\ve \frac{df}{d \ve}	 
         G^R_{\sigma}(\ve), 
\nonumber
\end{equation}
{\noindent}where $\Gamma = \frac{1}{2 \pi} |t|^2 \Delta$. 
In the regime of weak tunneling and $T > \Gamma$, 
it is possible to do a lowest order calculation in 
$|t|^2$ by using $G_R$ for an uncoupled dot, 
which can be obtained exactly \cite{Hackenbroich}.
Averaging over fluctuations of the dot wavefunctions,
and treating dot occupation factors 
in the same way as the rate-equations approach \cite{Beenakker}, we find
to lowest order in $|t|^2$, when $\Delta \ll T \ll E_C$,
\begin{eqnarray} \label{WeaktG}
    \frac{g}{g_{\rm max}} & = & 
	\frac{1}{2} - \cos(2 \delta) \left[
	\frac{1}{2} - \frac{|t|^2}{4} 
	\frac{E_C x / T}{ \sinh 2 E_C x / T } \right] \\
	&& + \sin(2 \delta) \left[
		\frac{|t|^2}{8} Y(x) \right]. 
	\nonumber
\end{eqnarray}
{\noindent}Near resonance $|x| \ll 1$, 
$Y(x)$ is approximately:
\begin{equation} \label{YForm}
    Y(x) = \frac{2}{\pi} \tanh(E_C x / T) \log \left[ \min \frac{2 E_C}{T},
    \frac{1}{|x|} \right].
\end{equation}
{\noindent}The first $|t|^2$ correction 
to the conductance in Eq. (\ref{WeaktG}) is
proportional to $\text{Im } G^R$,
and is identical to the conductance through a Coulomb
blockaded dot coupled to two leads when $\Delta \ll T
\ll E_C$ \cite{Beenakker}.  This term is always small ($\sim |t|^2
\ll 1$). In contrast, the second correction term to the
conductance, arising from $\text{Re } G^R$, gives rise
to a low-temperature logarithmic divergence near resonance. 
Its origin is a partial
cancellation between electron-like processes ($\ndot=N_0 \ra
N_0+1$) and hole-like processes ($\ndot = N_0 + 1 \ra N_0$),
which is identical to how a logarithmic divergence arises in
the conventional Kondo problem.  This is not surprising, given the
analogy already discussed.  The logarithm in 
Eq. (\ref{YForm}) is cut off by $x$,
consistent with $x$ playing the role of 
a magnetic field in the Kondo analogy.

Summarizing, we see that the lowest order in $|t|$
conductance calculation indicates an instability which enhances
the $\NN$-dependence of $g$ at low temperatures.  As
the only way an $\NN$-dependence can arise in the T-junction
geometry is via interference, this breakdown suggests an
enhancement of coherent scattering from the dot.  In terms of the
Kondo analogy, the failure of perturbation theory results from the
instability of the weak-coupling fixed point. The effective Kondo
temperature which characterizes this instability is $T_K \sim E_C
e^{ - c / |t|}$ (with $c$ a constant)\cite{MatveevKondo}, 
consistent with the fact that $|t|$ is
analogous to a dimensionless exchange coupling.

To investigate the regime $T, E_C x \leq T_K$, where Eq. 
(\ref{WeaktG}) fails and Kondo physics becomes dominant,
we now present results of a calculation for the opposite situation
of a strongly coupled quantum dot ($|t| \simeq 1$).  For strong
coupling, $T_K \ra E_C$,
and the condition $T < E_C$ ensures that we will be in a regime
dominated by Kondo physics for all values of $\NN$.
To deal with a strongly coupled dot at $\Delta \ll T \ll E_C$, we
use the approach of Flensberg \cite{Flensberg} and
Matveev \cite{Matveev1D} \cite {AGBReview} \cite{Schoeller}. 
In this approach the $\Delta \ra 0$ limit is
taken, and electron dynamics near the T-junction is described
using a one-dimensional model for each point contact. 
The interaction is treated
exactly using bosonization, while the effects of weak
backscattering ($|r| \ll 1$) are dealt with
perturbatively \cite{AashThesis}.  In what follows, we discuss
the case of spinless electrons and electrons with spin separately.

Without spin \cite{SpinlessNote}, 
our system corresponds to the single-channel Kondo model,
which is well known to have a Fermi liquid (FL) ground state in
which the magnetic impurity acts as a potential scatterer.  We
thus expect the $T < T_K \sim E_C$ properties of the open dot
system to also conform to a Fermi liquid state.  Indeed, a
rigorous calculation gives to order $|r|^2$:
\begin{equation} \label{SpinlessG}
    g = g_{\rm max} \sin^2 \left( \delta + \pi \ndot \right)
        + O \left( T/E_C \right)^2,
\end{equation}
where
\begin{eqnarray} 
    \ndot & = & \NN - \left( e^{\CC} |r| / \pi \right) \sin 2 \pi \NN
        \nonumber \\ 
	&& + \lambda_2 \left( e^{\CC} |r| / \pi  \right)^2
           \sin 4 \pi \NN + O(|r|)^3. \label{NdSpinless}
\end{eqnarray}
Here, $\CC$ is Euler's constant, and $\lambda_2 \simeq 1.9$. Eq.
(\ref{SpinlessG}) indicates that despite being at $T \gg \Delta$,
a regime where the non-interacting system is essentially
incoherent, reflection from the interacting dot is fully
coherent-- as $\NN$ is tuned, $g$ exhibits full
constructive and destructive interference.  The scattering phase
shift $\alpha = \pi \ndot$ obeys the Friedel sum rule
\cite{Friedel}, as expected for the FL ground state of the
single-channel Kondo model.  Equation (\ref{SpinlessG}) confirms 
that in the spinless
case, the breakdown of perturbation theory in
Eq. (\ref{WeaktG}) indeed signals coherent scattering from the dot.  The
fact that a Fermi liquid picture holds in the spinless case (near
$|r|=0$) was first noted by Aleiner and Glazman \cite{AG}.

\begin{figure}[t]
\centerline{\psfig{figure=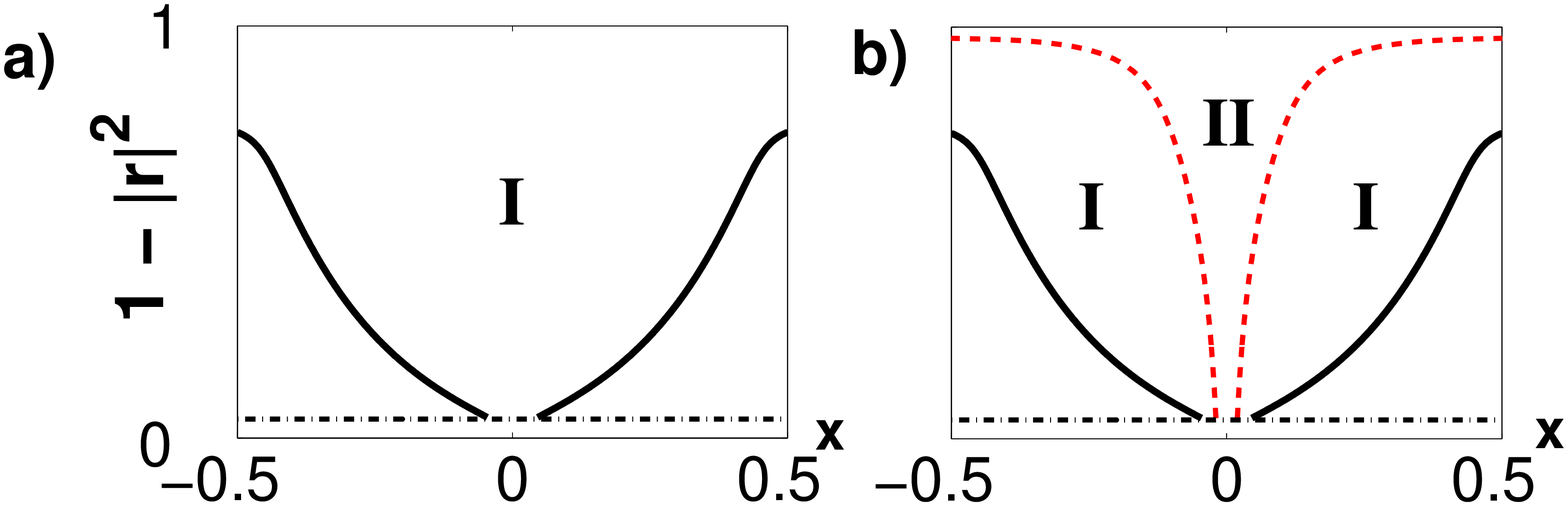,width=8.5cm}} 
\vspace{0.2 cm}
\caption{Heuristic phase diagrams for the conductance $g$ of 
the T-junction, for a fixed 
$T \ll E_C$, without spin (a) and with spin (b). 
$|r|^2$ is the reflection probability from the dot
entrance, and $x$ is the dimensionless gate voltage ($x=0$ implies 
charge degeneracy), c.f. Eq.(\ref{NDefn}).  
In regions I and II, infrared Kondo fixed points determine the physics;
the solid line indicates $E_C x = T_K(|r|)$, and the
dashed line in b) indicates $E_C x = \Gamma_c(|r|)$,
see text.  Scattering
from the dot is mainly coherent in region I (implying $g$ depends strongly 
on $x$ here), whereas it is mainly incoherent in II 
and in the unlabeled region outside I.
The dot-dashed line indicates $T = T_K(|r|)$. }
\vspace{-0.25 cm}
\label{SpinPhases}
\end{figure}

A heuristic phase diagram describing the $\NN$-dependence of the
conductance for a fixed temperature $\Delta \ll T < E_C$ is given in Fig.
\ref{SpinPhases}a.  For small $|r|$, $T < T_K$ for all $\NN$, and
the coherent expression of Eq. (\ref{SpinlessG}) holds for all $\NN$
(i.e., one is always in region I).  At $|r| \simeq 0$, 
$\ndot \simeq \NN$, and $g(\NN)$ has a sinusoidal form,
while for larger $|r|$, $\ndot$ 
will change rapidly by $1$ near resonance
\cite{Matveev1D}, implying a corresponding rapid change in the
phase $\alpha$ by $\pi$, and hence, a
Fano-type lineshape \cite{Fano}.
As we approach the weak coupling regime ($|r| \ra 1$), we
will have $T_K < E_C$, meaning that this Kondo induced coherence
will only occur at sufficiently low temperatures $T<T_K$ and 
close to resonance,  $|x| < T_K / E_C$. 
We still expect a narrow Fano lineshape in this
regime, as all the interesting phase behavior occurs near
resonance. For $T \geq T_K$, temperature cuts off scaling to the
strong-coupling fixed point, and consequently there will be no
enhancement of coherent scattering-- the $\NN$ dependence of $G$
will remain weak, being described by Eq. (\ref{WeaktG}).

Even though we are at $T \gg \Delta$,
including spin changes the behavior of the stub considerably. The
analogy is now to a two-channel Kondo model (the two spin
projections act as the two conserved channels), which is markedly
different from the the one-channel case.  At zero magnetic
field (i.e. $x=0$ in our system), the low temperature properties
of the 2CK model are described by a non-Fermi liquid (NFL) fixed
point which corresponds to a dimensionless exchange constant of
order unity (i.e., $|t| \simeq 1$).  A non-zero magnetic field (i.e.,
$|x| > 0$) destroys the stability of this fixed point, and the
system flows towards an alternate FL fixed point. 
Both these fixed points have an impact on the
conductance, as we now demonstrate.

Perturbation theory in $|r|$ for $T \ll E_C \sim T_K$ yields the
following form for the conductance:

\begin{equation} \label{FullSpinG}
   g = \frac{g_{\rm max}}{2} \Big(
    1 -
    \chi \cos(2\delta + 2 \alpha )  \Big),
\end{equation}
{\noindent}where, defining $\Gamma_{c} = 
2 e^{\CC} \pi^{-2} |r|^2 E_C \sin^2 (\pi x) $,
\begin{eqnarray}
    \chi(x,T) & = & c_1 \sqrt{\frac{\Gamma_{c}}{T}}
    + O \left(\frac{\Gamma_{c}}{T} \right)^{3/2},
    \label{ChiDefn} \\
    \alpha(x,T) & = & \frac{\pi}{2} \left(\frac{1}{2} + x -
    \theta(x) \right) +O(|r|)^2.
    \label{SpinPhaseDefn}
\end{eqnarray}
{\noindent}Here $c_{1} \simeq 1.8 $, and we have used Eq. 
(\ref{NDefn}) for the gate voltage $\NN$; it follows that 
$g$ is periodic in $\NN$ with period $1$.  The order $|r|^2$ 
correction to $\alpha$ is proportional to $\sin 2 \pi x$, and
diverges only logarithmically at low $T$.

The first term of Eq. (\ref{FullSpinG}) can be interpreted as an
incoherent contribution to the conductance, while the second term
represents an interference contribution.   At $|r|=0$ only the
former contributes, thus agreeing with what was found for the
non-interacting system, but in stark contrast to the
interacting spinless case, c.f. Eq. (\ref{SpinlessG}).  The
complete incoherence at $|r|=0$ corresponds to the vanishing
probability for single-particle scattering at the NFL fixed point
of the 2CK model\cite{Ludwig}. Equivalently, one can think of
fluctuations in the dot spin as suppressing a coherent response in
this temperature regime\cite{Konig}.
For non-zero $|r|$, the second term in Eq. (\ref{FullSpinG}) also
contributes.  This term has an interference form, with a
well-defined, $x$-dependent scattering phase $\alpha$ associated
with the dot.  The
weight $\chi$ of this term is zero on resonance ($x=0$), and
grows at low temperatures when off-resonance ($x \neq 0$). These
features can be understood within the 2CK analogy.  A non-zero $x$
makes the NFL fixed point
unstable, and the resulting renormalization group flow at low
temperatures is towards the FL fixed point, which has a
well-defined phase shift for scattering from the dot
\cite{Affleck}. This flow manifests itself here as a small
coherent term in the conductance which grows at low temperatures.
The flow is parameterized in Eq. (\ref{FullSpinG}) by the
function $\chi(x,T)$; we expect $\chi \ra 1$ in the vicinity of
the FL fixed point \cite{AGBReview}.

Note that the scattering phase $\alpha$ in Eq.
(\ref{SpinPhaseDefn}) is not simply proportional to $\ndot$. (At
$|r|=0$, $\ndot = \NN$ \cite{Matveev1D}). 
Near the FL fixed point (i.e. $T \ll
\Gamma_c(x)$), $\alpha$ can be obtained within the Kondo analogy
using Fermi liquid arguments \cite{Nozieres}. First, note that
$\left( \ndot - 1/2 - N_0 \right)$ is equivalent to $\left<S_z
\right>$, the moment of the Kondo impurity spin.  This moment will be
equal to the bare moment of the impurity plus a quasiparticle
contribution, which may be written in terms of phase shifts:
$\left<S_z \right> = \frac{1}{2} \sgn(x) + 2 \times \frac{1}{2}
(\frac{\delta_{\ua}(x)}{\pi} - \frac{\delta_{\da}(x)}\pi)$.  The
factor of two corresponds to the two equivalent channels of the
model. Finally, as the impurity spin has zero charge, one has 
$\delta_{\ua} = -
\delta_{\da}$.  In the Kondo analogy, $\ua$ is associated with
electrons in the lead.  Equating $\alpha$ with $\delta_{\ua}$ then
yields:
\begin{equation} \label{2CKalpha}
    \alpha = \frac{\pi}{2} \left( \ndot - N_0 - \theta(x) \right)
	\hspace{0.4 cm} \left[ \text{mod } \pi \right],
\end{equation}
which agrees with Eq. (\ref{SpinPhaseDefn}) \cite{WeakPhase}.

A heuristic phase diagram is shown in Fig. \ref{SpinPhases}b for a
fixed temperature $\Delta \ll T \ll E_C$.  For small
$|r|$, $\Gamma_c(x) < T$ for all $x$, and one
is always in region II--  the incoherent term in Eq. (\ref{FullSpinG}) 
dominates, and $g(x)$ exhibits only weak
oscillations.  These will grow as $|r|$ is increased or as $T$ is lowered,
following Eq. (\ref{FullSpinG}).
For sufficiently large $|r|$ (or low $T$),
tuning $x$ can take one from region II to region I, 
with the crossover occurring at $\Gamma_c(x) = T$ (dashed line
in Fig. \ref{SpinPhases}b).  Near resonance (in II), $g$ 
is still given by Eq.
(\ref{FullSpinG}), but away from resonance (in I), it is given by
the coherent expression $g = g_{\rm max} \sin^2(\delta + \alpha)$,
where $\alpha$ is given by Eq. (\ref{2CKalpha}). We expect large
oscillations in $g(\NN)$ in this regime, with a sharp feature
emerging around $x = 0$ as $T$ is lowered due to the rapid jump by
$\pi/2$ in $\alpha$.

The coherence effects discussed 
here are expected to be
largely insensitive to additional sources of dephasing in the 
quantum dot (e.g., from external sources or from 
electron-electron interaction terms we neglect) 
if $|t|$ (and hence $T_K$) is sufficiently large.  
For strong coupling ($|t| \ra 1$), the time an electron 
effectively spends in the dot 
before being reflected is $\sim \hbar / E_C$, which is  
much shorter than typical dephasing times \cite{MarcusReview}.  
Thus, electrons should still scatter coherently from the dot in this regime.  
Note that the model of Ref. \cite{Matveev1D} used for the strongly-coupled dot 
already assumes that electron motion in the dot is completely incoherent.

Finally, our results may have relevance to the
experiments in Ref. \cite{YacobyEtAl}, as they indicate
that the relation between $\ndot$ and 
the scattering phase from an interacting dot may be significantly
different than that expected for a non-interacting dot.
We thank L. I. Glazman, C. M. Marcus, K. Matveev, 
X. Waintal and X. G. Wen for discussions.  This work was supported by
the NSF under grants DMR-0086509 and DMR-9805613, 
by the Sloan Foundation, by the Olin Foundation, 
and by the Cornell Center for Materials Research.



\begin{references}
\bibitem{AlphaDefn}
    	We have 
	$e^{2 i \alpha} = e^{2 i \alpha_0} 
	\left(1 + |r| e^{-2 i \alpha_0} \right) / 
	\left(1 + |r| e^{2 i \alpha_0} \right)$,
    	where $\alpha_0$ is the 
	scattering phase of the dot when $|r| = 0$.
\bibitem{SDefn}
	Using the labeling in Fig. \ref{stubfig}a, 
	one has $|r|^2 = |S_{33}|^2 \equiv 1 - |t|^2$, 
	$g_{\rm max} = 8(|S_{31} S_{32}|^{2}) / |t|^4$, 
	and $\delta = \arg(S_{31} S_{32})$.  
	For spinless electrons, 
	$g_{\rm max} = 4(|S_{31} S_{32}|^{2}) / |t|^4$,
\bibitem{SpinlessNote}
  The case of spinless electrons can be realized experimentally
  by applying a large in-plane magnetic field to the
  T-junction, allowing the transmission of only one spin direction 
  into the dot.
\bibitem{WeakPhase}
	Note that the weak-coupling result of Eq. (\ref{WeaktG})
	can be written in the form of Eq. (\ref{FullSpinG}), with
	the phase $\alpha(\NN)$ being given again by Eq.
	(\ref{2CKalpha}).
\end{references}
\end{document}